\documentclass[pra,twocolumn,showpacs,10pt]{revtex4-1}

\usepackage{graphicx}

\usepackage{amsmath} 

\newcommand{\ket}[1]{\ensuremath{|#1\rangle}}
\newcommand{\braket}[2]{\langle#1|#2\rangle}

\newcommand{\ketup}{\ensuremath{|\mkern-5mu\uparrow\rangle}}
\newcommand{\ketupD}{\ensuremath{|\mkern-5mu\uparrow\varphi_2\rangle}}
\newcommand{\ketupT}{\ensuremath{|\mkern-5mu\uparrow\varphi_3\rangle}}
\newcommand{\ketdown}{\ensuremath{|\mkern-5mu\downarrow\rangle}}
\newcommand{\ketdownD}{\ensuremath{|\mkern-5mu\downarrow\varphi_2\rangle}}
\newcommand{\ketdownT}{\ensuremath{|\mkern-5mu\downarrow\varphi_3\rangle}}

\newcommand{\dd}[2]{ \frac{\partial #1}{\partial #2} }
\newcommand{\wf}{ \varphi }

\newcommand{\TLmatrix}[2]{\begin{array}{ccc} 0 & #1 & 0 \\ #1 & 0 &  #2 \\ 0 &  #2 & 0 \end{array}}
\newcommand{\refeq}[1]{(\ref{#1})}

\begin{document}

\title{Atomtronics with holes: \\ Coherent transport of an empty site in a triple well potential}
\author{A.~Benseny,$^{1}$ S.~Fern\'andez-Vidal,$^{1}$ J.~Bagud\`a,$^{1}$ R.~Corbal\'an,$^{1}$ A.~Pic\'on,$^{1,2}$ L.~Roso,$^{3}$ G.~Birkl,$^{4}$ and J.~Mompart$^{1}$}
\affiliation{$^{1}$Grup d'\`{O}ptica, Departament de F\'{i}sica, Universitat Aut\`{o}noma de Barcelona, E-08193 Bellaterra, Spain}
\affiliation{$^{2}$JILA, University of Colorado, Boulder 80309-0440, USA (present address)}
\affiliation{$^{3}$Centro de L\'aseres Pulsados (CLPU), E-37008 Salamanca, Spain}
\affiliation{$^{4}$Institut f\"ur Angewandte Physik, Technische Universit\"at Darmstadt, Schlossgartenstr. 7, D-64289 Darmstadt, Germany}
\date{\today}

\begin{abstract}
We investigate arrays of three traps with two fermionic or bosonic atoms. The tunneling interaction between neighboring sites is used to prepare multi-site dark states for the empty site, i.e., the hole, allowing for the coherent manipulation of its external degrees of freedom. By means of an \textit{ab initio} integration of the Schr\"odinger equation, we investigate the adiabatic transport of a hole between the two extreme traps of a triple-well potential.  Furthermore, a quantum-trajectory approach based on the de Broglie--Bohm formulation of quantum mechanics is used to get physical insight into the transport process. Finally, we discuss the use of the hole for the construction of a coherent single hole diode and a coherent single hole transistor.

\end{abstract}

\maketitle

\section{Introduction \label{sect:intro}}

The beginning of the 21st century has brought the development of techniques to isolate and manipulate individual neutral atoms \cite{Singleatom}, allowing to follow a bottom-top approach where quantum systems acquire classical features as their size and/or their coupling with the environment increases.
In fact, in the last few years a lot of attention has been devoted to the field of \textit{atomtronics} \cite{atomtronics}, where atomic matter waves in optical \cite{Bir1,Bir2,COLD}, magnetic \cite{Magnetictraps} and electric \cite{Electricpot} potentials play an analogous role to electrons in electronic devices.
Atomtronics has the important advantage over electronics that neutral atoms are comparatively less sensitive to decoherence than charged particles, i.e., interaction with the ``classical'' environment can be almost completely inhibited for the former. 
In fact, neutral atom devices based on the coherent tunneling of matter waves are, in general, designed to take profit of their inherent quantum features.
In this context, several proposals on diode- and transistor-like behaviors have been deeply investigated for ultracold atoms \cite{Muga07} and Bose--Einstein condensates \cite{Zoller04} both in double and triple well optical potentials as well as in optical lattices with applications ranging from atomic cooling to quantum information processing.

Although atomtronic devices have yet to be realized experimentally, the time for coherent atomtronics is already here mainly due to the fact that techniques for cooling and trapping atoms are by now very well established \cite{Singleatom,Bir1,Bir2,COLD,Magnetictraps}.
Neutral atoms can be stored and manipulated in optical lattices, standard dipole traps, and microtraps.
In particular, magnetic and optical microtraps offer an interesting perspective for storing and manipulating arrays of atoms with the eventual possibility to scale, parallelize, and miniaturize the atomtronic devices.
Moreover, optical microtraps can take advantage of the fact that most of the current techniques used in atom optics and laser cooling are based on the optical manipulation of atoms.
In fact, the possibility to store and to selectively address single optical microtraps, as well as initializing and reading out the quantum states in each of the sites has been already reported \cite{Bir1,Bir2}.

In this context, there is a need for the development of novel techniques to control the coherent flow of matter waves in optical and magnetic traps based on tunneling devices.
Recently, we introduced a set of coherent tools \cite{TLAO1}, named three-level atom optics (TLAO) techniques, to efficiently transport matter waves between the two extreme traps of a triple-well potential via the tunneling interaction.
This adiabatic transport process is the matter wave analog of the very well known quantum optical STIRAP technique \cite{STIRAP} and it is based on adiabatically following an energy eigenstate of the system, the so-called spatial dark state, that, ideally, only involves the vibrational ground states of the two extreme wells.
Extensions of these TLAO techniques to atomic wave packets propagating in dipole waveguides \cite{TLAO2}, to Bose--Einstein condensates \cite{matterwave_STIRAPs}, to the transport of electrons in quantum dot systems \cite{dots}, and to superconductors \cite{superconductors}, have been later performed.
Even very recently, by exploiting the wave analogies between classical and quantum systems, de Longhi \textit{et al.} \cite{Longhi} have experimentally reported, for the first time, light transfer in an engineered triple-well optical waveguide  by means of the classical analog of the matter wave STIRAP.

In the first part of this paper we will consider an array of three traps with two neutral atoms and extend the TLAO techniques to the transport of the empty site: the hole, see Fig.~\ref{fig:kai}(a).
In the second part of the paper, we will discuss the use of the hole as an active player in coherent {\it atomtronic} devices \cite{atomtronics}.
We will design a single hole diode, see Fig.~\ref{fig:kai}(b), by tuning the interaction between the atoms, allowing for the hole transport to be successful in one direction but inhibited in the opposite.
Furthermore, we will engineer a single hole transistor, see Fig.~\ref{fig:kai}(c), by addressing and manipulating the spin of an individual atom and changing the symmetry of the two atoms spin state. 

The article is organized as follows.
In Section \ref{sect:pm} we will introduce the physical system under investigation: an array of three traps with two neutral atoms.
In Section \ref{sect:transport} we will describe the adiabatic transport of a hole between the two extreme traps introducing the concept of a spatial dark state for the hole. Through a numerical integration of the Schr\"odinger equation, we will address the adiabatic dynamics of a single hole in a triple well potential. 
By means of a quantum-trajectory approach based on the de Broglie-Bohm formulation of quantum mechanics, we will elucidate some specific features of the adiabatic transport technique here proposed. 
This Section will be followed by a discussion, Section \ref{sect:holetronics}, on the use of the hole as a key element in the building up of a coherent single hole diode and a coherent single hole transistor. In Section \ref{sect:conc} the concluding remarks are presented.

\section{Physical System \label{sect:pm}}

\begin{figure}[thbp]
\includegraphics{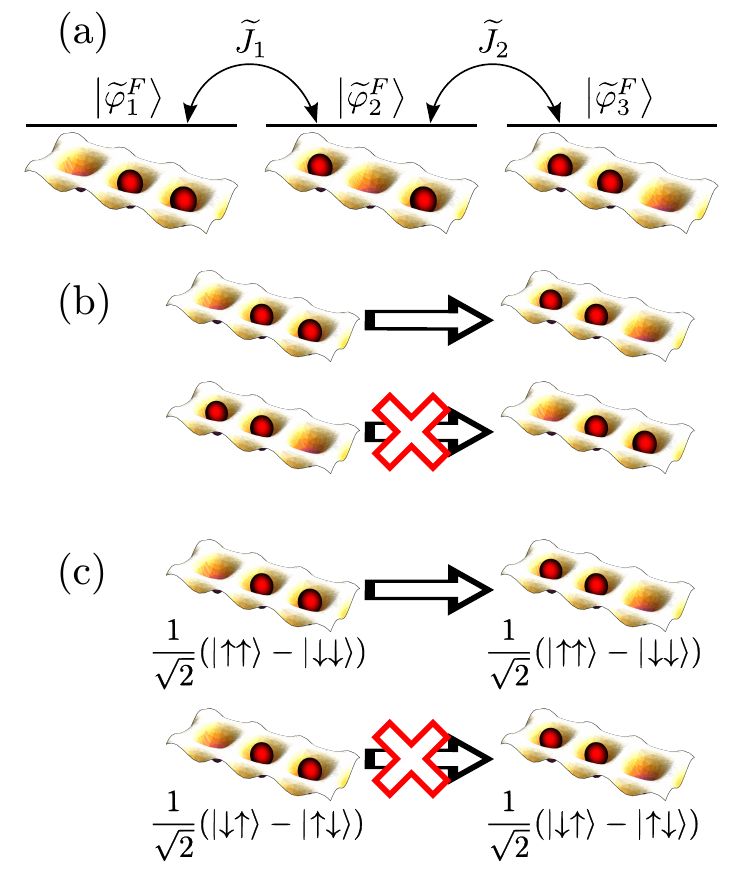}
\caption{(color online)
(a) Simplified three-level description of the system under investigation:
a three-trap array loaded with two atoms, e.g., two fermions, and one empty site, i.e., a hole.
Matter wave STIRAP techniques will be applied to transfer the hole between the traps.
$\ket{\widetilde{\wf }^F_i}$ is the localized state for the (fermionic) hole being at trap $i=1,2,3$ and $\widetilde{J}_i$ is the hole tunneling rate between traps $i$ and $i+1$.
(b) Single hole diode: for identical trap-approaching schemes but with appropriately tuned interactions between two bosonic atoms, the hole will be transported from the left to the right trap, but the inverse process will not succeed.
(c) Single hole transistor for two non-interacting fermionic atoms: by performing identical trap-approaching schemes, the hole transport from left to right will succeed or will be inhibited depending on whether the spin state of the two atoms is symmetric or antisymmetric.}
\label{fig:kai}
\end{figure}

The system under investigation is sketched in Fig.~\ref{fig:kai}(a) and consists of an array of three optical traps that has been loaded with two  fermionic or bosonic atoms in the lowest vibrational levels.
For the physical implementation of the dipole traps, we will consider the 2D array of optical microtraps discussed in Refs. \cite{Bir1,Bir2} where single site addressing as well as the ability to approach columns (or, alternatively, rows) of traps yielding coherent atomic transport has been demonstrated \cite{Bir2}.
Note that the use of the 2D array allows for the performance of the techniques here described in a parallel fashion, realizing multiple experiments simultaneously.

We assume that tunneling between sites occurs only in the column movement direction (namely, $x$) and thus, the main dynamics will be restricted to one dimension. For two identical atoms of mass $m$, the dynamics of the system are governed by the Hamiltonian:
\begin{eqnarray}
H &=& \sum_{i=1,2} \left[ -\frac{\hbar^2}{2m} \dd{^2}{x_i^2} + V(x_i) \right] + U(x_1, x_2)  ,\label{Hamones}
\end{eqnarray}
with $V$ being the trapping potential and $U$ the interaction between the two atoms.

For simplicity, we assume that the three-trap array potential consists of truncated harmonic wells centered at positions $\left\{ x_{0_i} \right\}$ with $i = 1$, 2, 3:
\begin{eqnarray}
V(x) = \frac{1}{2} m \omega_x^2 \, \min_{i} \left\{ (x- x_{0_i})^2 \right\}   ,\label{eq-V}
\end{eqnarray}
where $\omega_x$ is the longitudinal trapping frequency of each trap. For the one-dimensional model to be valid, we assume a tight transverse confinement such that transverse excitations can be neglected, i.e., $\omega_p \gg \omega_x$, being $\omega_p$ the transverse trapping frequency.
In this one-dimensional model, the cold collisional interaction between the atoms can be modeled by a contact potential of the form \cite{Cal00}:
\begin{eqnarray}
U (x_1, x_2) = 2 \hbar a_s \omega_p \delta(x_1 - x_2)  , \label{eq-U}
\end{eqnarray}
with $a_s$ the $s$-wave scattering length.

\section{Adiabatic transport of holes \label{sect:transport}}

Following standard ideas of solid state physics we address the present problem in terms of a hole for either fermionic or bosonic atoms \cite{Roso}. 
For the hole description to be valid, the following conditions must be fulfilled: 
(i) each trap contains, at most, one atom;
(ii) all atoms are cooled down to the vibrational ground state of each trap; and
(iii) tunneling is adiabatically controlled to strongly suppress the probability of double occupancy.
To satisfy condition (iii) for identical fermions, we assume that the two atoms have parallel spins such that the Pauli exclusion principle applies and double occupancy in the same vibrational state is strictly forbidden. For bosons, we require for the $s$-wave scattering length to be large enough to, in the adiabatic limit, inhibit double occupancy.

For the two fermions case, the entire two-atom state must be antisymmetric which means that if their spin state is symmetric (antisymmetric) then the spatial wavefunction must be antisymmetric (symmetric).
For the two bosons case, since the entire two-atom state must be symmetric, then their spin state and their spatial wavefunction must have the same symmetry.
Taking into account that the dynamics we will simulate only involve the spatial wavefunction, we will distinguish between the two cases where:
(i) the spatial wavefunction is antisymmetric (fermions with symmetric spin state or bosons with antisymmetric spin state), and refer to it as the fermionic case; and
(ii) the spatial wavefunction is symmetric (bosons with symmetric spin state or fermions with antisymmetric spin state), and refer to it as the bosonic case.
In the fermionic case, the contact potential from Eq. \refeq{eq-U} will not play any role in the dynamics since at $x_1=x_2$ the (antisymmetric) spatial wavefunction vanishes.
Furthermore, the fermionic and hardcore (strongly interacting) bosonic cases will present equivalent dynamics \cite{hardcorebosons} due to the fermionic exchange interaction.

\begin{figure}[thbp]
\includegraphics[width=1.0\linewidth]{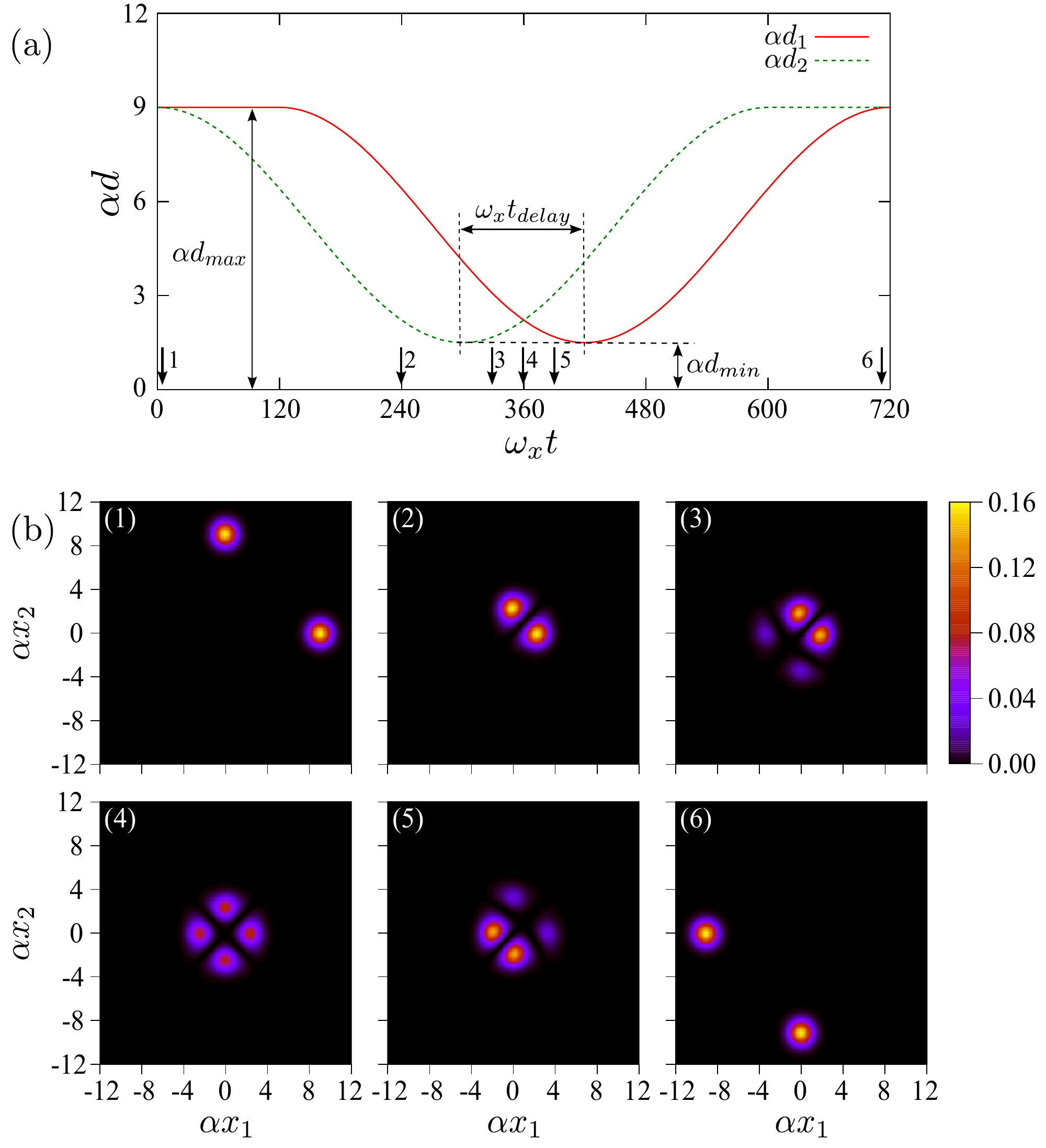}
\caption{(color online) Coherent transport of a fermionic hole in a triple well potential:
(a) Temporal variation of the trap distances in terms of $\alpha$ (its inverse is the width of the ground state of each isolated trap) and $\omega_x$ (the longitudinal trapping frequency). 
(b) Snapshots of the two-fermion joint probability distribution for the particular times indicated by the arrows in (a).
Initial state: $\phi_1^F(x_1, x_2) = \braket{x_1, x_2}{\widetilde{\wf}^F_1}$. Parameter values: $\alpha d_{max}=9$, $\alpha d_{min}=1.5$ and $\omega_x t_{delay} = 120$.}
\label{fig:dd} 
\end{figure}

Our first goal will consist in developing an efficient and robust method to adiabatically transport holes between the two extreme traps of the triple well potential, by applying the matter wave analogue \cite{TLAO1} of the quantum optical STIRAP technique \cite{STIRAP}.
Let us consider, thus, three in-line traps with one empty site and two identical fermions, each one in the vibrational ground state of the two remaining traps as shown in Fig.~\ref{fig:kai}(a) (the bosonic case will be discussed later on). In a three-state approximation, where the Hilbert space of the system is restricted to the lowest three energy eigenstates, the spatial wavefunction of the two atoms can be expressed in the following basis: 
\begin{eqnarray}
\ket{\widetilde{\wf }^F_1} \equiv \frac{1}{\sqrt{2}} \left[\ket{\wf_2 }_1 \ket{\wf_3}_2 - \ket{\wf_3}_1 \ket{\wf_2}_2 \right]   , \label{statshole1}\\
\ket{\widetilde{\wf }^F_2} \equiv \frac{1}{\sqrt{2}} \left[\ket{\wf_3 }_1 \ket{\wf_1}_2 - \ket{\wf_1}_1 \ket{\wf_3}_2 \right]   , \label{statshole2}\\
\ket{\widetilde{\wf }^F_3} \equiv \frac{1}{\sqrt{2}} \left[\ket{\wf_1 }_1 \ket{\wf_2}_2 - \ket{\wf_2}_1 \ket{\wf_1}_2 \right]   , \label{statshole3}
\end{eqnarray}
where $\ket{\wf_j}_k = \ket{\wf_j (x,t)}_k$ is the time-dependent state of the $k$-th atom localized in the $j$-th trap. States $\ket{\widetilde{\wf}^F_i }=\ket{\widetilde{\wf}^F_i (x_1,x_2,t)}$ with $i=1,2,3$ account for the fermionic hole being in the left, middle, and right trap, respectively.  Note that Eqs.~(\ref{statshole1}--\ref{statshole3}) are antisymmetric since we have assumed that the spin state of the atoms is symmetric.
For truncated harmonic traps, see Eq.~\refeq{eq-V}, the $J_i$ tunneling rate of a single atom between the ground states of two adjacent traps $i$ and $i+1$ is given by \cite{TLAO1}:
\begin{eqnarray} 
J_i(\alpha d_i) = \omega_x \frac{ -1 +e^{(\alpha d_i)^2}\left[ 1+\alpha d_i\left( 1-\mathop{\rm erf}(\alpha d_i)\right) \right]}{\sqrt{\pi }\left( e^{2 (\alpha d_i)^2}-1\right)/ (2 \alpha d_i) }   , \label{J_tunnel}
\end{eqnarray}
where $d_i \equiv \left| x_{0_{i+1}} - x_{0_i} \right|$ and $\alpha \equiv \sqrt{m\omega_x /\hbar}$. Therefore, in the hole basis $\left\{ \ket{\widetilde{\wf}^F_1}, \, \ket{\widetilde{\wf}^F_2}, \, \ket{\widetilde{\wf}^F_3} \right\}$ the dynamics of the system are governed by the Hamiltonian: 
\begin{eqnarray}
H_{3\, {\rm TRAPS}} = \hbar \left( \TLmatrix{\widetilde{J}_1(t)}{\widetilde{J}_2(t)} \right)  , \label{HamJJ}
\end{eqnarray}
being $\widetilde{J}_i(=J_i)$ \cite{JtildeJ} the hole tunneling rate between two adjacent traps, see Fig.~\ref{fig:kai}(a). One of the three eigenstates of Hamiltonian \refeq{HamJJ} is the so-called spatial dark state \cite{TLAO1} that only involves the two states where the hole is in the extreme traps:
\begin{equation} \label{DarkState}
\ket{\widetilde{D}^F(\Theta(t))} = \cos \Theta(t) \, \ket{\widetilde{\wf}^F_1} - \sin \Theta(t) \, \ket{\widetilde{\wf}^F_3}  ,
\end{equation}
with $\tan \Theta(t) = {\widetilde{J}_1(t)} / {\widetilde{J}_2(t)}$. The transport of the hole between the two extreme traps of the triple well potential consists in  adiabatically following state $\ket{\widetilde{D}}$ from $\ket{\widetilde{\wf}^F_1}$  to $\ket{\widetilde{\wf}^F_3}$ by smoothly varying the mixing angle $\Theta $ from $\Theta = 0 $ to $\Theta = \pi /2$. 
As in standard optical STIRAP \cite{STIRAP}, it is convenient to establish a general adiabaticity criterion given by
$\widetilde{J}_{\rm max} t_{\rm delay } > A$, 
with $\widetilde{J}_{\rm max}^2 \equiv (\widetilde{J}_1)_{\rm max}^2+(\widetilde{J}_2)_{\rm max}^2$ and $A$ being a dimensionless constant that for optimized temporal delays and tunneling profiles takes values around 10.
The generalization of Hamiltonian \refeq{HamJJ} and the adiabatic transport process to a trap array of arbitrary length can be found in Appendix A.

The previously discussed three-level approach has been introduced to illustrate the main ideas behind the hole transport. Nevertheless, and in order to be accurate, in what follows we will numerically solve the Schr\"odinger equation in real space. 
Figs.~\ref{fig:dd}(a) and \ref{fig:dd}(b) show an exact simulation, i.e., an \textit{ab initio} numerical integration of the Schr\"odinger equation with the Hamiltonian given in Eq.~\refeq{Hamones}, of the adiabatic transport process of a single hole in a triple well potential with two identical fermions.
The initial state is $\phi^F_1(x_1, x_2) = \braket{x_1, x_2}{\widetilde{\wf}^F_1}$ with $\alpha d_1=\alpha d_2=9$ while the expected final state is, up to a global phase, $\phi^F_3 (x_1, x_2)= \braket{x_1, x_2}{\widetilde{\wf}^F_3}$ with $\alpha d_1=\alpha d_2=9$. 
For the truncated harmonic potentials here considered, at $\alpha d_i = 9$ the tunneling rate between adjacent traps is almost negligible and they can be considered as isolated.
For the time variation of the trapping potential, we have taken the middle trap to be static at $x=0$ while displacing only the two extreme traps. Note that the hole transport sequence, Fig.~\ref{fig:dd}(a), starts by first approaching the two occupied traps and later approaching the empty trap to the middle one.
Fig.~\ref{fig:dd}(b) shows different snapshots for the temporal evolution of the two-fermion joint probability distribution $\left| \phi(x_1, x_2, t) \right|^2$. Note that the diagonal $x_1 = x_2$ is forbidden due to the Pauli principle and the probability density is mirrored at both sides of this diagonal due to the antisymmetrization of the wavefunction.

As indicated in Eq.~\refeq{DarkState}, the hole is transferred from the left to the right trap with an ideally negligible population in the middle one.
Therefore, the signature that the hole has been transferred through the matter wave STIRAP technique is that the counter-diagonal $x_1 = - x_2$ is practically not populated (see the 3rd, 4th, and 5th snapshots of Fig.~\ref{fig:dd}(b)). 
However, resorting to the continuity equation associated with the two-atom matter wave, the corresponding wavefunction must cross at some point the {\it forbidden} counter-diagonal. To get physical insight into this particular feature of the adiabatic transport process, we will discuss now very briefly the previous simulations by means of quantum trajectories \textit{\'a la de Broglie-Bohm} \cite{Bohm}. See Appendix B for details of the quantum trajectories formulation.

\begin{figure}[thbp]
\includegraphics[width=0.99\linewidth]{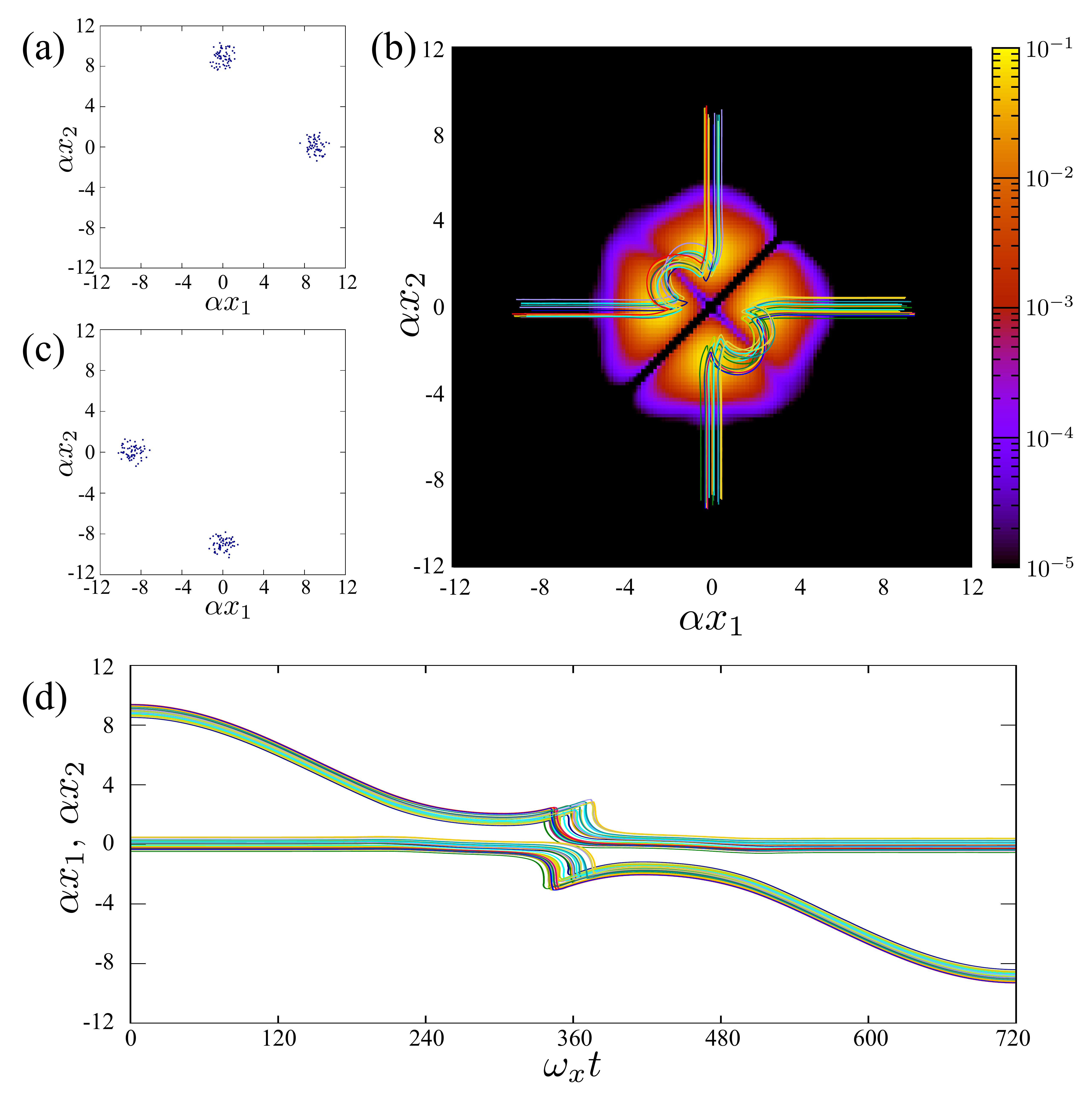}
\caption{(color online) de Broglie-Bohm trajectories corresponding to the temporal evolution of the system shown in Fig.~\ref{fig:dd}(b). (a) Initial distribution of quantum trajectories (cf. Fig.~\ref{fig:dd}(b-1)). (b) Evolution of the trajectories in configuration space together with the joint probability distribution at $\omega_x t = 360$. (c) Final distribution of trajectories (cf. Fig.~\ref{fig:dd}(b-6)). (d) Evolution of the trajectories as a function of time.
In order to allow for the easy visualization of the transport process, (b) and (d) only show a reduced number of quantum trajectories. }
\label{fig:trajs}
\end{figure}

Fig.~\ref{fig:trajs} shows a set of quantum trajectories calculated from the time evolution of Fig.~\ref{fig:dd}(b). Their initial positions, see Fig.~\ref{fig:trajs}(a), are randomly distributed according to $\left|\phi^F_1(x_1, x_2) \right|^2$. As expected, Fig.~\ref{fig:trajs}(b) reveals that the time evolution of the quantum trajectories follows the evolution of the wavefunction, ending up distributed according to $\left|\phi^F_3(x_1, x_2)\right|^2$, see Fig.~\ref{fig:trajs}(c). In Fig.~\ref{fig:trajs}(b) we have also plotted the atomic probability distribution for the intermediate time $\omega_x t=360$ corresponding to the 4th snapshot in Fig.~\ref{fig:dd}(b). 
Clearly, when crossing the forbidden counter-diagonal, each quantum trajectory suddenly increases its velocity, see Fig. \ref{fig:trajs}(d), such that the density of trajectories per unit time vanishes in this counter-diagonal. In addition, quantum trajectories make a detour from the central region of the counter-diagonal where the probability distribution is significantly smaller.

\begin{figure}[thbp]
\includegraphics{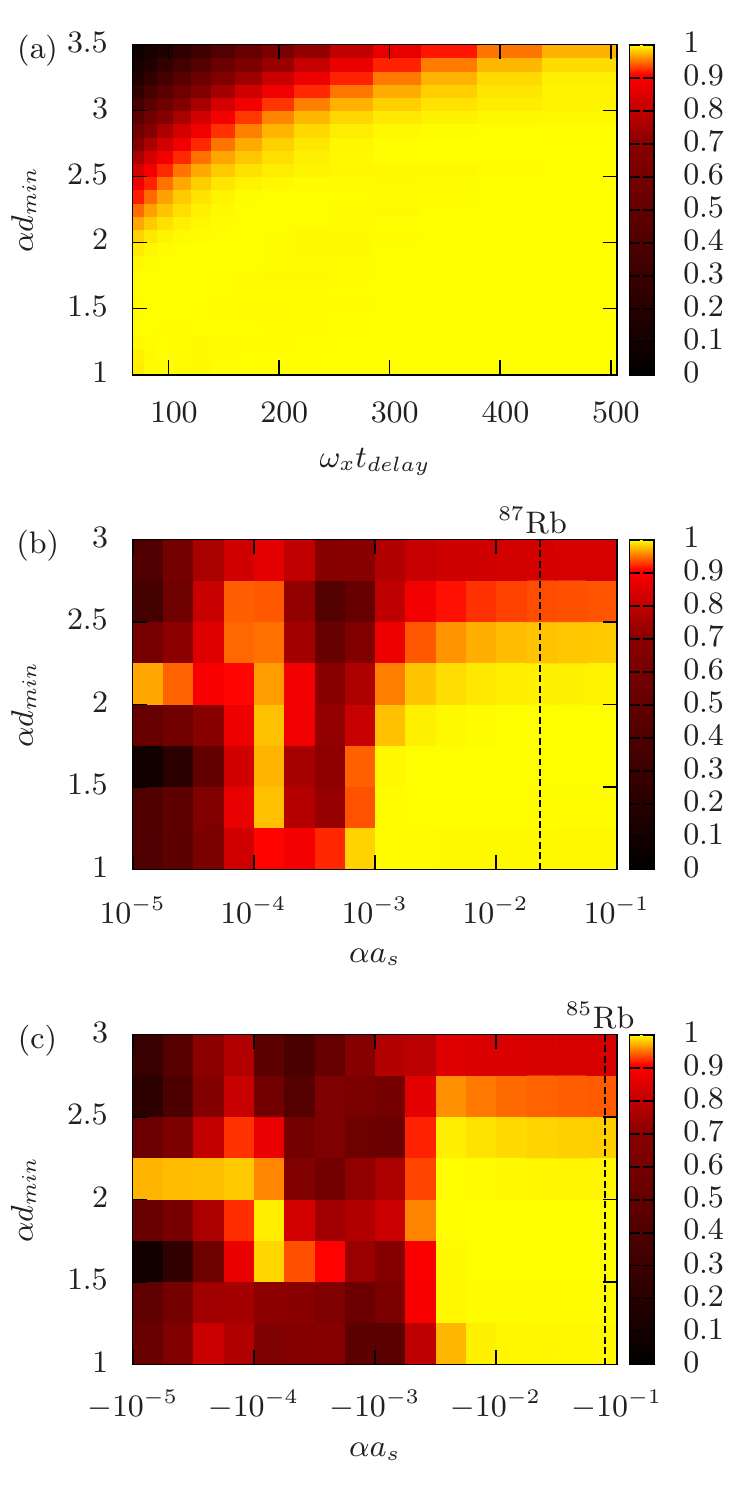}
\caption{(color online) Matter wave STIRAP fidelity: (a) $F_{1 \rightarrow 3}^F$ in the parameter plane $\{t_{delay},d_{min}\}$. (b) $F_{1 \rightarrow 3}^B$ in the parameter plane $\{a_{s},d_{min}\}$ for $a_s$ positive and (c) for $a_s$ negative.
Scattering lengths of $^{85}$Rb ($\alpha a_s = -7.98 \cdot 10^{-2}$) and $^{87}$Rb ($\alpha a_s = 2.32 \cdot 10^{-2}$) are indicated by dashed black lines. The temporal variation of the traps and the rest of the parameters is given in Fig.~\ref{fig:dd}(a).
Transverse trapping frequency $\omega_p = 24 \omega_x$ \cite{Bir2} and, for the bosonic case, $\omega_x t_{delay} = 120$.
The bright (yellow) area indicates the area where the fidelity is larger than 0.99.}
\label{fig:fids}
\end{figure}

For the characterization of the transport probabilities between the different traps, after applying the matter wave STIRAP sequence shown in Fig.~\ref{fig:dd}(a), we define here their associated fidelities.
These definitions will be also very useful for the characterization of the atomtronic devices introduced in Sec. \ref{sect:holetronics}.

Thus, by starting with the hole in trap $i=1,2,3$ and performing the temporal evolution, we check the population of each trap $j=1,2,3$.
We denote, for either a fermionic ($F$) or a bosonic ($B$) hole case, the state of the system at the end of the process as $\phi_i^{F/B}(x_1, x_2, T)$, being $T$ the total duration of the STIRAP sequence.
The population in the localized state $\phi_j^{F/B}(x_1,x_2)$ after this evolution will be given by the product between this state and the evolved one, namely:
\begin{equation}
F_{i \rightarrow j}^{F/B} = \left| \iint  {\phi_j^{F/B}}^*(x_1, x_2)\phi_i^{F/B}(x_1, x_2, T)  dx_1 dx_2 \right|^2 .
\end{equation}
Thus, $F_{i \rightarrow j}^{F/B}$ is the fidelity of the fermionic ($F$) or bosonic ($B$) transport process of the hole between traps $i$ and $j$ after applying the hole matter wave STIRAP sequence, see Fig.~\ref{fig:dd}(a).
Note that for the matter wave adiabatic transport process, we want to maximize $F_{1 \rightarrow 3}^{F/B}$.

Fig.~\ref{fig:fids}(a) depicts $F_{1 \rightarrow 3}^F$ in the parameter plane $\{t_{delay},d_{min}\}$ for the hole transport process (see Fig.~\ref{fig:dd}(a) for the definition of these two parameters).
It comes clear that for a large set of parameters the fidelity is larger than 0.99 (see the bright yellow area in Fig.~\ref{fig:fids}(a)), showing that the hole transport from left to right via matter wave STIRAP is a robust and efficient technique, provided that the adiabaticity condition is fulfilled.

We have also simulated the hole transfer process for the case of two bosonic atoms by integrating the corresponding Schr\"odinger equation.
In this case, the localized states for the bosonic hole, \ket{\widetilde{\wf }^B_i}, are given by the symmetrized versions of Eqs.~(\ref{statshole1}--\ref{statshole3}).
Figs.~\ref{fig:fids}(b-c) show $F_{1 \rightarrow 3}^B$ in the parameter plane $\{a_{s}, d_{min}\}$. 
As shown in the figures, the adiabatic transfer process succeeds for $\alpha a_s = -7.98 \cdot 10^{-2}$ and $\alpha a_s = 2.32 \cdot 10^{-2}$ corresponding, respectively, to the $s$-wave scattering length of $^{85}$Rb and $^{87}$Rb \cite{Bir1}, while it breaks down for weaker interactions since then double occupancy starts to play a dominant role.
For large absolute values of the $s$-wave scattering length, bosons become hardcore and then their dynamics is equivalent to that of the fermionic case.

\section{Atomtronics with holes \label{sect:holetronics}}

Making use of the fact that the hole transfer process here presented is spatially non-symmetric, we will now discuss both a coherent single hole diode and a coherent single hole transistor in a triple-well potential. The hole transfer from left to right and vice versa strongly depends on both the two-atom collisional interaction and the exchange interaction and, therefore, both interactions will be used here to control the diode and transistor operation regimes.

\subsection{Single hole diode}

\begin{figure}[htbp]
\includegraphics{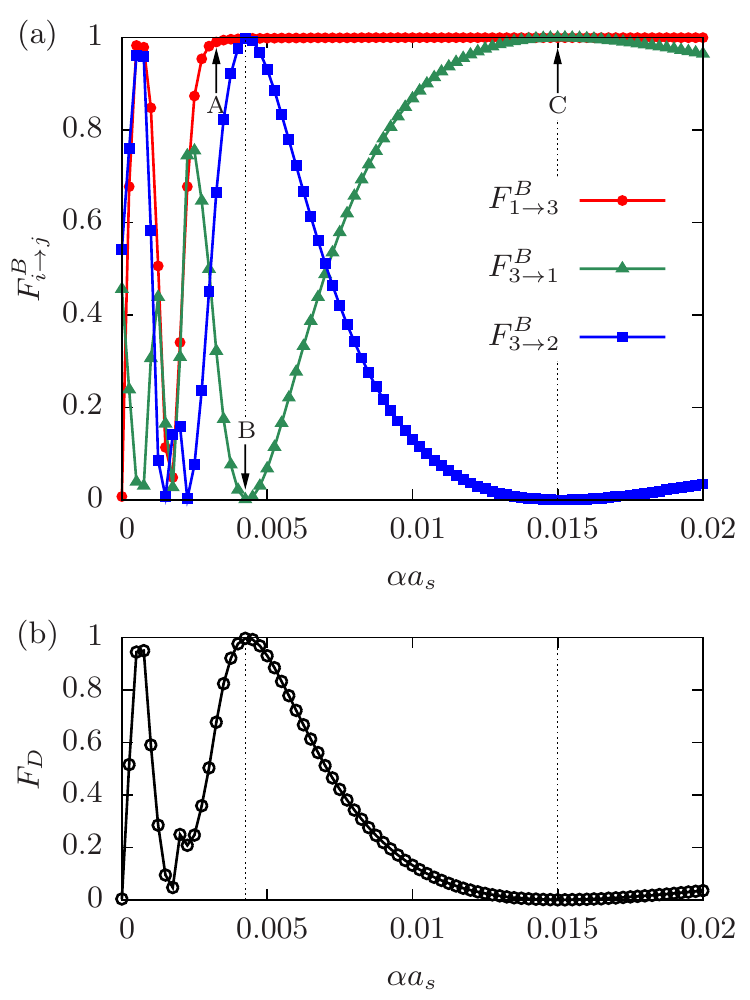}
\caption{(color online) Single hole diode: For a bosonic hole and as a function of the $s$-wave scattering length: (a) Fidelities of hole transport processes $F_{1 \rightarrow 3}^B$ (red circles), $F_{3 \rightarrow 1}^B$ (green triangles) and $F_{3 \rightarrow 2}^B$ (blue squares) and (b) diode fidelity $F_D$ (see Eq.~\refeq{eq:diode_F} in the text). See text for the definitions of points A, B and C.
The rest of the parameter values are as in Fig.~\ref{fig:dd}.}
\label{fig:diode}
\end{figure}

In this section we will design a single hole diode by using the collisional interaction between two bosons as a control parameter to allow the hole transport from left to right, and inhibit the transport from right to left, see Fig.~\ref{fig:kai}(b).
Thus, Fig.~\ref{fig:diode}(a) shows the fidelity of the bosonic hole transport processes $F_{1 \rightarrow 3}^B$, $F_{3 \rightarrow 1}^B$ and $F_{3 \rightarrow 2}^B$ against the strength of the $s$-wave scattering length.
The parameter values for the temporal variation of the traps are taken as in Fig.~\ref{fig:dd} such that the fidelity of the hole transport process from left to right, $F_{1 \rightarrow 3}^B$ (red circles in Fig.~\ref{fig:diode}(a)), is larger than 0.99 above a certain threshold value for the scattering length, indicated by point A in Fig.~\ref{fig:diode}(a), i.e., when the interaction is strong enough and the bosons become hardcore.

By performing the same trap-approaching scheme but with the hole starting on the right trap, the process corresponding to the hole being transferred from the right to the left trap (green triangles in Fig.~\ref{fig:diode}(a)) is inhibited ($F_{3 \rightarrow 1}^B \sim 0$ at point B) or succeeds ($F_{3 \rightarrow 1}^B \sim 1$ at point C), depending on the value of $\alpha a_s$.

We thus define the fidelity of the diode as:
\begin{equation}
F_D = F_{1 \rightarrow 3}^B \left( 1 - F_{3 \rightarrow 1}^B\right)  . \label{eq:diode_F}
\end{equation}
since this fidelity is maximal when the bosonic hole is transported from left to right and, simultaneously, the opposite process consisting in the hole transport from right to left is inhibited, i.e., $F_D=1$ when $F_{1 \rightarrow 3}^B = 1$ and $F_{3 \rightarrow 1}^B = 0$.

As can be seen in Fig.~\ref{fig:diode}(b), that shows $F_D$ as a function of the scattering length, by tuning the product of the inverse of the size of the wavefunction and the $s$-wave scattering length to $\alpha a_s \sim 4.25\cdot10^{-3}$, where $F_D \sim 1$ (corresponding to point B), we obtain a scheme that transports the hole from the left to the right trap but transfers a hole from the right trap to the middle one (see blue squares in Fig.~\ref{fig:diode}(a)).
Note that an ideal diodic behavior where the hole ends at the right trap, no matter if initially it was at the left or the right trap, would violate the unitarity of the quantum evolution. 

\subsection{Single hole transistor}

Fig.~\ref{fig:fids}(a) shows that in the fermionic case, the hole transport from left to right achieves high fidelities.
On the other hand, Figs.~\ref{fig:fids}(b-c) show that in the weakly interacting bosonic case, i.e., for $a_s \rightarrow 0$, the hole transport does not perfectly succeed.
In fact, we have checked that for $a_s = 0$ the fidelity $F_{1 \rightarrow 3}^B$ vanishes with the hole ending in a superposition between being in the left and middle traps.
From the previous observations, it is possible to design a single hole transistor where the spin state of the atoms is used to control the hole current from the left to the right trap.

For instance, it is straightforward to check that, for two fermions in the middle and right traps (hole in the left), the state with symmetric spin state $\ket{S}=(\ketup_1\ketup_2 - \ketdown_1\ketdown_2)/\sqrt{2}$ and antisymmetric spatial state (\ket{\widetilde{\wf }^F_1}, cf. Eq. \refeq{statshole1}),
\begin{eqnarray}
\ket{S} \ket{\widetilde{\wf }^F_1} =  \frac{1}{2} [&&\ketupD_1\ketupT_2 - \ketupT_1\ketupD_2 \nonumber \\* &&- \ketdownD_1\ketdownT_2 + \ketdownT_1\ketdownD_2] ,
\end{eqnarray}
and the state with antisymmetric spin state $\ket{A}=(\ketdown_1\ketup_2 - \ketup_1\ketdown_2)/\sqrt{2}$ and symmetric spatial state (\ket{\widetilde{\wf }^B_1}),
\begin{eqnarray}
\ket{A} \ket{\widetilde{\wf }^B_1} =  \frac{1}{2} [&& \ketdownD_1\ketupT_2 + \ketdownT_1\ketupD_2 \nonumber \\* &&-\ketupD_1\ketdownT_2 - \ketupT_1\ketdownD_2] ,
\end{eqnarray}
are coupled via a spin flip on the atom in the middle trap, i.e., $\ketupD_k \leftrightarrow \ketdownD_k$. A similar argument for bosonic atoms can be done between states $\ket{S}\ket{\widetilde{\wf }^B_1}$ and $\ket{A}\ket{\widetilde{\wf }^F_1}$.

This control over the system behavior between the bosonic and fermionic cases allows us to create a coherent hole transistor scheme where the matter wave STIRAP sequence from the left to the right trap succeeds or is inhibited depending on the spin state of the atoms.
The case for two non-interacting fermions is depicted in Fig.~\ref{fig:kai}(c).
Thus, the figure of merit corresponds to the maximization of the transistor fidelity defined as:
\begin{equation}
F_T = F_{1 \rightarrow 3}^F \left( 1 - F_{1 \rightarrow 3}^B\right)  . \label{eq:transistor_F}
\end{equation}
since it will be maximal when $F_{1 \rightarrow 3}^F \sim 1$ and $F_{1 \rightarrow 3}^B \sim 0$.
As we have discussed, for parameters of Fig.~\ref{fig:dd}(a), $F_T > 0.99$ for non-interacting atoms.

\begin{figure}[htbp]
\includegraphics{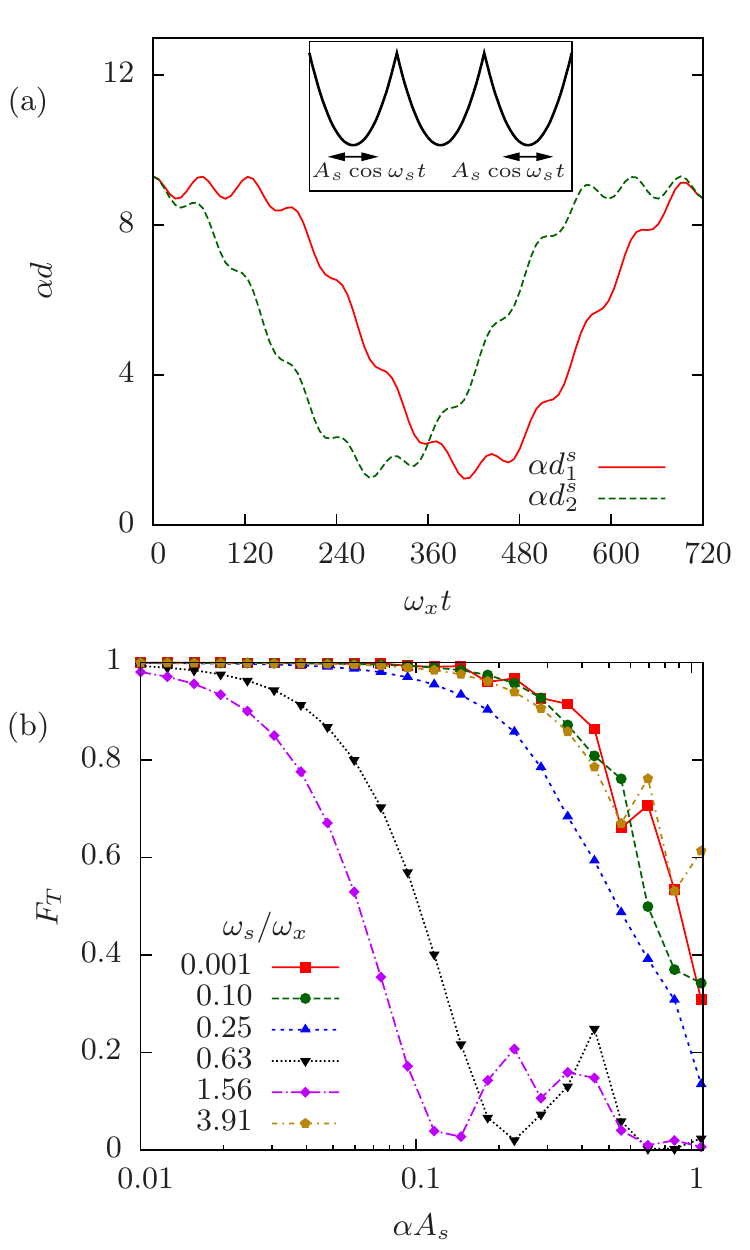}
\caption{(color online) Single hole transistor:
(a) Temporal variation of the trap distances shown in Fig. \ref{fig:dd}(a) with an added jittering of amplitude $\alpha A_s = 0.3$ and frequency $\omega_s = 0.1 \omega_x$. The inset shows a sketch of the three traps and its simulated jittering.
(b) Fidelity $F_T$ of the transistor (see Eq.~\refeq{eq:transistor_F}) as a function of the amplitude of position jitter for different jitter frequencies.
$a_s = 0$, and the rest of the parameter values are as in Fig. \ref{fig:dd}.}
\label{fig:transistor} 
\end{figure}

To further test the fidelity of the atomtronic transistor, we have calculated $F_T$, computing $F_{1 \rightarrow 3}^{F/B}$, adding a jitter in the trap positions (to simulate experimental imperfections), see Fig.~\ref{fig:transistor}(a), such as:
\begin{equation}
d_i^s(t) = d_i^0(t) + A_s \cos \omega_s t \label{eq:shake}  ,
\end{equation}
being $d_i^0(t)$ the distance between the $i$-th and $i+1$-th trap shown in Fig.~\ref{fig:dd}(a). $A_s$ and $\omega_s$ represent, respectively, the amplitude and the frequency of the jitter.
The results of $F_T$ for different values of $A_s$ and $\omega_s$ are plotted in Fig.~\ref{fig:transistor}(b) and indicate that for small jitter amplitudes the transistor still works with high fidelity ($F_T>0.99$) for a wide range of frequencies, except for those close to $\omega_x$ (trapping frequency) that, as expected, excite the atoms to unwanted vibrational states.

\section{Conclusions \label{sect:conc}}

Using the matter wave analogue \cite{TLAO1} of the quantum optical STIRAP technique \cite{STIRAP}, we have proposed an efficient and robust method to coherently transport empty sites, i.e., holes, in arrays of three dipole traps with two neutral atoms.
The coherent transport process consists in adiabatically following a spatially delocalized dark state by an appropriate temporal control of the tunneling rates.
We have first introduced the transport process in a simplified three-state model to, later on, simulate it with exact numerical integrations of the Schr\"odinger equation for time dependent potentials.
Some particular features of the adiabatic matter wave dynamics, such as the transport through forbidden regions, have been elucidated by means of de Broglie-Bohm quantum trajectories.
Finally, making use of both the collisional interaction and the exchange interaction, we have analyzed in detail hole transport schemes for the implementation of a coherent single hole diode and a coherent single hole transistor in a triple well potential with two neutral atoms. 

Additionally, we want to note that it is possible to engineer a quantum information processing scheme that uses the hole as the qubit carrier, the computational states defined by the presence of the hole in one out of the two extreme traps of a triple well potential. 
The implementation of single qubit gates could be performed by taking advantage of the tunneling between traps, and a controlled phase gate between two qubits could be implemented by using the collisional interaction between the atoms, in a similar manner as described in Ref. \cite{SDQ}.
In fact, the hole transport process discussed along the paper could be used to coherently prepare particular qubit states for the hole.

\begin{acknowledgments}

We would like to thank Ver\`{o}nica Ahufinger, Gabriele De Chiara, and Xavier Oriols for fruitful discussions.
We acknowledge support from the Spanish Ministry of Education and Science 
under contracts FIS2008-02425 and CSD2006-00019, the Catalan Government under contract SGR2009-00347, and the DAAD through the program Acciones Integradas Hispano-Alemanas HD2008-0078.
A. B. acknowledges financial support through grant AP 2008–01275 from MICINN (Spain).

\end{acknowledgments}

\appendix

\section{Single hole in an array of $n$ (odd) traps}

In most quantum computation proposals with trapped neutral atoms, a defect-free quantum system where all sites of the lattice are occupied by exactly one atom is needed to start the information processing, e.g., empty sites must be removed from the physical area of computation.
With this purpose, here we will extend the adiabatic transport method presented in the main text to a trap array of arbitrary length by means of the Hubbard model, generalizing Hamiltonian \refeq{HamJJ} to a system of $n$ (odd) traps.
This technique could be used to transport holes away from the area of interest in order to prepare defect-free trap domains to, eventually, perform quantum computations.

Note that the exact numerical simulation, i.e., by means of the corresponding Schr\"odinger equation, of the adiabatic transport of a hole in an array with a large number of single-occupancy traps is extremely demanding from a computational point of view and out of the scope of the present paper. However, the Hubbard formulation for the hole that we will now introduce could be used to simulate the hole dynamics following the lines of previous works on multi-level STIRAP \cite{multirap}.

Let us consider a single-occupancy 1D array of $n$ (odd) traps loaded with $n-1$ spin-polarized fermions (or hardcore bosons) that, therefore, presents one isolated defect consisting of an empty site in one of its extremes. Again, the goal is to adiabatically transfer this empty site, i.e., the hole, from one extreme of the array to the other. 
The dynamics of a spin polarized non interacting Fermi gas loaded in a 1D trap array is described by the fermionic Hubbard Hamiltonian \cite{Wandev}:
\begin{equation}
\hat{H}  = -\hbar \sum_{i} J_{i} \left(\hat{c}^{\dag}_{i} \hat{c}_{i+1} + \hat{c}^{\dag}_{i+1} \hat{c}_{i}\right). \label{ham2}
\end{equation}
Operators $\hat{c}^{\dag}_{i}$ and $\hat{c}_{i}$ are the fermionic creation and annihilation operators at site $i$ satisfying fermionic anticommutation relations and $J_i$ accounts for the tunneling between neighboring sites (see Eq.~\refeq{J_tunnel}).
We have dropped the on-site energy term \cite{Wandev} since we are considering a homogeneous system with fixed number of atoms.

In this context, we consider a hole (an empty site) as a virtual particle whose vacuum state, \ket{\widetilde{\Omega}}, corresponds to all sites occupied with one fermion, reading:
\begin{equation}
\ket{\widetilde{\Omega}} \equiv \hat{c}^{\dag}_{1}\hat{c}^{\dag}_{2}\ldots \hat{c}^{\dag}_{n}\ket{\Omega}  ,
\end{equation}
with $\ket{\Omega}$ the fermionic vacuum state. Since we do not allow for transitions to excited vibrational states and consider $n$ traps with $n-1$ atoms, the dynamics of our system is constrained to remain in states with a single hole, $\{ \hat{C}_i^{\dag} \ket{\widetilde{\Omega}} \}$, with $\hat{C}_i^{\dag} = \hat{c}_i$ being the hole creation operator at site $i$.
Then, in terms of these on-site hole operators, and hole tunneling rates $\widetilde{J}_i(=J_i)$, Hamiltonian \refeq{ham2} reads:
\begin{eqnarray}
\hat{H} = -\hbar \sum _{i} \widetilde{J}_{i}\left(\hat{C}^{\dag}_{i+1}\hat{C}_{i}+\hat{C}^{\dag}_{i}\hat{C}_{i+1}\right)  . \label{hamhole}
\end{eqnarray}
For $n=3$ we retrieve Hamiltonian \refeq{HamJJ}. For $n$ odd, it is straightforward to check that this Hamiltonian has an energy eigenstate:
\begin{eqnarray}
\ket{\widetilde{D}}=\sum^{\frac{n+1}{2}}_{m=1}\left(-1\right)^{m+1}\left(\prod^{m-1}_{j=1}\widetilde{J}_{2m-2j-1}\right) \times  \nonumber \\
\left(\prod^{\frac{n-1}{2}-m}_{j=0}\widetilde{J}_{2m+2j}\right)\hat{C}^{\dag}_{2m-1}\ket{\widetilde{\Omega}}  , \label{f-ds}
\end{eqnarray}
that satisfies $\hat{H}\ket{\widetilde{D}} = 0$. Note that $\ket{\widetilde{D}}$ not only involves the first and last traps, but also all $\hat{C}_i^\dag \ket{\widetilde{\Omega}}$ with $i$ odd. In this case, the hole transfer following state $\ket{\widetilde{D}}$ would be achieved by favoring first the tunneling rates $\widetilde{J}_i$ with even $i$ and then the ones with odd $i$ \cite{multirap,multirap-Australians}. Finally, notice also that hole transport based on the adiabatic following of the multi-site spatial dark state given in \refeq{f-ds} can be straightforwardly applied to the case of hardcore bosons.

\section{de Broglie-Bohm formulation}

By casting the polar form of the wavefunction, $\wf=Re^{i S / \hbar}$, into the Schr\"{o}dinger equation (with the conventional meaning of the symbols),
\begin{eqnarray}
i \hbar \frac{\partial \wf}{\partial t} = - \frac{\hbar^2}{2m} \frac{ \partial^2 \wf}{\partial x_1^2} - \frac{\hbar^2}{2m} \frac{ \partial^2 \wf}{\partial x_2^2} + V \wf , \label{app:schro}
\end{eqnarray}
and separating real and imaginary parts, we obtain \cite{Bohm}
\begin{eqnarray}
- \frac{\partial S}{\partial t} &=& V
+ \sum_{i=1,2} \frac{1}{2m} \left( \frac{\partial S}{\partial x_i}  \right)^2 \nonumber \\* & & - \sum_{i=1,2} \frac{\hbar^2}{2 m} \frac{1}{R} \frac{\partial^2 R}{\partial x_i^2} , \label{app:QHJeq}  \\
- \frac{\partial R^2}{\partial t} &=& \sum_{i=1,2}  \frac{1}{m} \frac{\partial}{\partial x_i}  \left( R^2 \frac{\partial S}{\partial x_i} \right) . \label{app:QConteq}
\end{eqnarray}
Eq.~\refeq{app:QHJeq} is the so-called quantum Hamilton--Jacobi equation because of its similarity with the (classical) Hamilton--Jacobi equation but with one additional term, the quantum potential, accounting for the quantum features of the system. This similarity suggests the definition of the particle velocity as:
\begin{eqnarray}
v_i[t] &=& \left.\frac{1}{m} \frac{\partial S}{\partial x_i} \right|_{(x_1[t], x_2[t], t)}. \label{app:bohmvel}  
\end{eqnarray}
Thus, Eq.~\refeq{app:QConteq} becomes a continuity equation ensuring that the trajectories distribution is given by $R^2(x_1, x_2, t)$ at all times. After solving Eq. \refeq{app:schro} (or, alternatively, Eqs.~\refeq{app:QHJeq} and \refeq{app:QConteq}) and distributing the initial positions of trajectories following the probability density function $R^2(x_1, x_2, t_0)$, we find the quantum trajectories (time evolution of the positions) as:
\begin{eqnarray}
x_i[t] &=& \int_{t_0}^t v_i[t] dt .
\end{eqnarray}
We have followed the previous approach to obtain the quantum trajectories displayed in Fig.~\ref{fig:trajs}.

\bibliographystyle{apsrmp}
\bibliography{CMES}

\begin{references}
\bibitem{JtildeJ}
These tunneling rates are equal since the atom and hole transport processes between two adjacent traps are complementary.
\end{references}

\end{document}